\documentclass[aps,twocolumn,showpacs]{revtex4}
 \usepackage{epsfig,rotating,minitoc}
\usepackage{amsmath,amssymb,amsthm}
\usepackage{graphicx}
\usepackage{psfrag}
\usepackage{bm}
\usepackage[dvips]{color}

\theoremstyle{definition}

\theoremstyle{remark}

\begin{document}

\title{Parity-symmetry-adapted coherent states and entanglement in quantum phase
  transitions of vibron models}

\author{M. Calixto}
\affiliation{Departamento de Matem\'atica Aplicada, Universidad de Granada,
Campus de Fuentenueva S/N, 18071 Granada, Spain. E-mail: calixto@ugr.es}
\author{E. Romera}
\affiliation{Departamento de F\'{\i}sica At\'omica, Molecular y Nuclear and
Instituto Carlos I de F{\'\i}sica Te\'orica y
Computacional, Universidad de Granada, Campus de Fuentenueva S/N, 18071 Granada,
Spain. E-mail: eromera@ugr.es}
\author{R. del Real}
\affiliation{Departamento de F\'{\i}sica At\'omica, Molecular y Nuclear, Universidad de Granada, Fuentenueva s/n, 18071 Granada,
Spain. E-mail: ram.delreal@gmail.com}

\date{\today}
\begin{abstract}
We propose coherent (`Schr\"odinger catlike') states adapted to the parity symmetry providing a remarkable variational description of
the ground and first excited states of vibron models for finite-($N$)-size molecules.
Vibron models undergo a quantum shape phase transition (from linear to bent) at a critical value $\xi_c$ of a control parameter.
These trial cat states reveal a sudden increase of vibration-rotation entanglement linear ($L$) and von Neumann ($S$) entropies from zero to 
$L^{(N)}_{\rm cat}(\xi)\simeq 1-{2}/{\sqrt{\pi N}}$
[to be compared with $L^{(N)}_{\rm max.}(\xi)=1-{1}/{(N+1)}$] and $S^{(N)}_{\rm cat}(\xi)\simeq \frac{1}{2} \log_2(N+1)$, respectively, 
above the critical point, $\xi>\xi_c$, in agreement with exact numerical calculations.
We also compute inverse participation ratios, for which
these cat states capture a sudden delocalization  of the ground
state wave packet across the critical point. Analytic expressions for
entanglement entropies and inverse participation ratios of variational states, as functions of $N$ and $\xi$,
are given in terms of hypergeometric functions.

\end{abstract}
\pacs{03.65.Fd, 33.20.Vq, 05.30.Rt, 03.65.Ud}


\maketitle

\section{Introduction}
 Understanding (zero temperature) quantum phase transitions (QPTs) \cite{sachdev} has become an important
part of quantum many-body theory. On the one hand, quantum fluctuations grow at the QPT and variance and entropic uncertainty
measures turn out to give a good description of the QPT. Recently  \cite{nuestro1,nuestro2,nuestro3}, uncertainty measures
in the (spin-boson) Dicke model \cite{dicke}, which also undergoes a QPT,  have
been  studied  (see also \cite{pla,pha,jsm}
for other information theoretical measures of delocalization to study QPTs). On the other hand, large correlations
and collective behavior are an intrinsic part of critical systems and, therefore, entanglement measures
should also capture the essence of this QPTs (see e.g. \cite{entang1,entang2} for entanglement properties
of the Dicke model and \cite{entang3} for entanglement and localization of a two-mode Bose-Einstein
condensate).

In this article we shall focus on the study of entanglement properties of the so-called `vibron models' \cite{muchos},
which are used to study rotational and vibrational spectra in
diatomic and polyatomic molecules and also exhibit a (shape) QPT. They were introduced  by Iachello \cite{Iachello1}
in the 80's through an algebraic approach, based on the concept of spectrum-generating algebra.
We shall restrict ourselves to the two-dimensional $U(3)$ vibron model describing a system
containing a dipole degree of freedom constrained to planar motion \cite{Iachello2,Iachello3}.
The basic example of such a system is a triatomic linear bender molecule, although
extensions to more complex molecular systems can also be considered. In these models,
one finds different (shape) phases connected to specific geometric configurations of
the ground state and related to distinct dynamic symmetries of the Hamiltonian
(see e.g. \cite{curro}). The  QPT occurs
as a function of a control parameter $\xi$ that appears in the Hamiltonian $H$ in
the form of a convex combination $H(\xi)=(1-\xi)H_1+\xi H_2$. At $\xi=0$ the system
is in phase I, characterized by the dynamical symmetry $G_1$ of $H_1$, and at $\xi=1$
the system is in phase II, characterized by the dynamical symmetry $G_2$ of
$H_2$.

The classical,
thermodynamic or mean-field, limit of these models is studied by using an algorithm introduced by Gilmore
\cite{Gilmore} which makes use of semi-classical
(boson-condensate) Coherent States (CSs) (see e.g. \cite{Perelomov,Klauder,Gazeau,Vourdas} for  standard references on CSs), as variational states
to approximate the ground state energy. However, these trial states, as such,
do not display an intrinsic parity symmetry (see e.g. \cite{parityop}) present in the Hamiltonian, as
they do not have a definite parity. We would like to mention that the key role of parity has also been recently noticed by
\cite{casta1,casta2}  in the context of QPT (from normal to superradiant)  in the Dicke model for
matter-field (spin-boson) interactions. We shall see that, disregarding parity, leads to wrong results of vibration-rotation entanglement measures for usual 
variational approximations to the ground state in terms of (non-symmetry-adapted) CSs. Then  we shall introduce even (resp. odd) CSs  adapted to the parity symmetry
of the Hamiltonian to describe the ground (resp. first-excited) state of the two-dimensional $U(3)$ vibron model. These even and odd CSs
are ``Schrodinger's catlike-states'' in the sense that they are a quantum
superposition of quasi-classical, macroscopically distinguishable (with negligible overlap) states. Schrodinger's cat states are experimentally generated
(see e.g. \cite{gato}) and prove to be
very useful to study the process of decoherence which limits the development of quantum computing.
We shall show that even-cat states provide finite-($N$)-size approximation to
some  $N=\infty$ quantities like the ground state energy
`per particle' and order parameters like the `equilibrium radius' (see also \cite{curro2} for other analytical results of
finite-size corrections beyond the mean-field
limit approximation in vibron models). Even-cat states will also quantitatively and qualitatively capture entanglement and delocalization measures of the 
exact ground state for finite $N$. We have to say that finite-size effects have also been discussed in the above-mentioned Dicke model, both numerically \cite{epl69} and
analytically \cite{epl74}, for thermodynamical quantities and in \cite{jsm07} for the entanglement entropy, although the role of parity is not explicit in these
studies (see \cite{casta1,casta2} and \cite{nuestro1,nuestro2,nuestro3} for the relevance of parity in
uncertainty measures inside the Dicke model). Finite-size corrections in \cite{epl74} and in \cite{prc} (for two-level boson systems) use a $1/N$ expansion naturally
given by the Holstein-Primakoff representation of the angular momentum \cite{HP}. For finite-size precursors of QPTs in atomic nuclei models see
for example \cite{rmp} and references therein. We must stress that, in this paper, we do not make any of these $1/N$ expansions around $N=\infty$, but we are working 
with finite $N$. Numerical (exact) results of the ground state energy, entanglement and delocalization measures are already nicely reproduced by 
even-cat states for finite $N$.

This article is organized as follows. In Sec. \ref{sec2} we  briefly review the $U(3)$ algebraic approach to
two-dimensional (2D) vibron models (more details can be found in \cite{Iachello2,Iachello3,curro}) and we  propose a new kind of variational
states by adapting `projective' (`Hartree axial' \cite{prc}) CSs to the parity symmetry of the Hamiltonian. In Sec. \ref{sec3} we compute vibration-rotation entanglement
measures like `purity', linear and von Neumann entropy, as a function of the vibron number $N$
(related to the total number of bound states of the molecule) and the control parameter $\xi$, revealing a sudden increase of
entanglement from linear to bent phases at the critical point $\xi_c$. The calculation is done numerically and complemented and 
compared with two variational approximations (parity-symmetric and
non-symmetric), which enrich the study. Finally, we compute the inverse participation ratio (IPR), as a measure of delocalization
of the ground state wave packet across the phase transition, and see that the spreading is captured by
the new proposed symmetry-adapted (even-cat) CSs, thus revealing the
importance of parity symmetry. Sec. \ref{sec4} is devoted to conclusions.

\section{Vibron model and variational symmetry-adapted coherent states\label{sec2}}
  2D-vibron models  describe
a system containing a dipole degree of freedom constrained to planar motion.
Elementary excitations are (creation and annihilation) 2D vector $\tau$-bosons $\{\tau_x^\dag, \tau_y^\dag, \tau_x,
\tau_y\}$ and a scalar $\sigma$-boson $\{\sigma^\dag,\sigma\}$. It is convenient to introduce circular bosons:
$\tau_\pm=\mp(\tau_x\mp i\tau_y)/\sqrt{2}$. The nine generators of the $U(3)$ algebra are bilinear products of
creation and annihilation operators, in particular:
\begin{eqnarray}
 &\hat{n}=\tau_+^\dag\tau_++\tau_-^\dag\tau_-,\, \hat{n}_s=\sigma^\dag\sigma,&\nonumber\\
&\hat{l}=\tau_+^\dag\tau_+-\tau_-^\dag\tau_-,&\\
&\hat{D}_+=\sqrt{2}(\tau^\dag_+\sigma-\sigma^\dag\tau_-),\; \hat{D}_-=\sqrt{2}(-\tau^\dag_-\sigma+\sigma^\dag\tau_+),&\nonumber
\end{eqnarray}
denote the number operator of vector $\hat n$ and scalar $\hat{n}_s$ bosons, 2D angular momentum $\hat{l}$ and dipole $\hat D_\pm$ operators, respectively
(see \cite{curro} for the reminder four operators $\hat{Q}_\pm,\hat{R}_\pm$, which will not be used here).  Assuming
the total number of bosons $\hat{N}=\hat{n}+\hat{n}_\sigma$ and the 2D angular momentum $\hat{l}$ to be conserved,
there are only two dynamical symmetry limits, $G_1=U(2)$ and $G_2=SO(3)$, associated with two algebraic
chains starting from $U(3)$ and ending in $SO(2)$:  the so-called `cylindrical' and `displaced' oscillator chains.  A general Hamiltonian of the $U(3)$ vibron model
with only one- and two-body interactions can be expressed in terms of linear and quadratic Casimir operators of all the
subalgebras contained in the dynamical symmetry algebra chains. To capture the essentials of the phase transition
from the $G_1$-phase (linear) to the $G_2$-phase (bent) it is enough to consider a convex combination of
the linear $C_1(U(2))=\hat{n}$ and quadratic $C_2(SO(3))=\hat{W}^2=(\hat{D}_+\hat{D}_-+\hat{D}_-\hat{D}_+)/2+\hat{l}^2$
Casimir operators of the corresponding dynamical symmetries. In particular, we shall consider the essential
Hamiltonian \cite{curro}
\begin{equation}
 \hat{H}=(1-\xi)\hat{n}+\xi\frac{N(N+1)-\hat{W}^2}{N-1},\label{hamiltonian}
\end{equation}
where the (constant) quantum number $N$ is the total number of bound states that labels the totally symmetric
$(N+1)(N+2)/2$ dimensional representation $[N]$ of $U(3)$. The  Hilbert space is spanned by the orthonormal basis vectors
\begin{equation}
 |N;n,l\rangle=\frac{(\sigma^\dag)^{N-n}(\tau^\dag_+)^{\frac{n+l}{2}}(\tau^\dag_-)^{\frac{n-l}{2}}}
{\sqrt{(N-n)!\left(\frac{n+l}{2}\right)!\left(\frac{n-l}{2}\right)!}}|0\rangle,\label{basis}
\end{equation}
where the bending quantum number $n=N,N-1,N-2,\dots,0$ and the angular momentum
$l=\pm n,\pm(n-2),\dots,\pm 1$ or $0$ ($n=$odd or even) are the eigenvalues of
$\hat{n}$ and $\hat{l}$, respectively. The matrix elements of $\hat{W}^2$ can be easily derived
(see e.g. \cite{curro}):
\begin{equation}
\begin{array}{l}
 \langle N;n',l|\hat{W}^2|N;n,l\rangle= \\  ((N-n)(n+2)+(N-n+1)n+l^2)\delta_{n',n}\\
 -((N-n+2)(N-n+1)(n+l)(n-l))^{\frac{1}{2}} \delta_{n',n-2}\\
 -((N-n)(N-n-1)(n+l+2)(n-l+2))^{\frac{1}{2}}\delta_{n',n+2}.
\end{array}\nonumber
\end{equation}
From these matrix elements, it is easy to see that time evolution preserves the parity $e^{i\pi n}$ of a given state
$|N;n,l\rangle$. That is, the parity operator $\hat\Pi=e^{i\pi \hat{n}}$ commutes with $\hat{H}$ and both operators
can then be  jointly diagonalized. Parity symmetry in the vibron model has been considered in,
for instance, Ref. \cite{parityop3}. However, this symmetry goes unnoticed when
proposing ansatzs to approximate the ground state energy in terms of
(non-symmetric) `projective' \cite{curro} (or `Hartree axial' \cite{prc}) CSs
\begin{equation}
 |N;r\rangle\equiv\frac{1}{\sqrt{N!}}(b_c^\dag)^N|0\rangle,\;\;  b^\dag_c=\frac{1}{\sqrt{1+r^2}}(\sigma^\dag+r\tau^\dag_x),\label{pcs}
\end{equation}
with $r\geq 0$ a free variational parameter and $b^\dag_c$
the boson condensate. Other rotationally equivalent possibilities can be also
considered \cite{bosonrot}; moreover, intrinsic excitations can also be constructed 
by replacing the intrinsic boson $b_c$ with
orthogonal excitation bosons, thus defining  multi-species CSs (see e.g \cite{kuyucak,caprio}). In this
article we shall restrict ourselves to ground state ansatzs.

For future use, we shall provide the explicit expression of the coefficients of the expansion of (\ref{pcs})
in terms of the basis vectors (\ref{basis}), which is  explicitly given by:
\begin{eqnarray}
|N;r\rangle&=&\sum_{n=0}^N\sum_{m=0}^n c_{n,m}^{(N)}(r)|N;n,n-2m\rangle,\label{csinbasis}\\
 c_{n,m}^{(N)}(r)&=&\sqrt{\binom{N}{n}\binom{n}{m}}\frac{(-r/\sqrt{2})^n(-1)^m}{{(1+r^2)}^{N/2}}\nonumber.
\end{eqnarray}
The variational parameter $r$ is fixed by minimizing the ground state energy
functional `per particle' \cite{curro}:
\begin{eqnarray}
 {\cal E}_\xi(r)&=&\frac{\langle\hat{H}\rangle}{N}=(1-\xi)\frac{\langle\hat{n}\rangle}{N}+\xi\frac{N(N+1)-
\langle\hat{W}^2\rangle}{N(N-1)}\nonumber\\
&=&
(1-\xi)\frac{r^2}{1+r^2}+\xi
\left(\frac{1-r^2}{1+r^2}\right)^2,\label{energyns}
\end{eqnarray}
where we have used $\langle\cdot\rangle$ as a  shorthand for expectation values in $|N;r\rangle$. From $\partial {\cal E}_\xi(r)/\partial r=0$ one gets
the `equilibrium radius' $r_e$ and the ground state energy $ {\cal E}_\xi$
as a function of the control parameter $\xi$:
\begin{equation}
 \begin{array}{l}
r_e(\xi)=\left\{\begin{array}{ll} 0, &  \xi\leq \xi_c=1/5\\
\sqrt{\frac{5\xi-1}{3\xi+1}}, &\xi> \xi_c=1/5 \end{array}\right.\\
 {\cal E}_\xi(r_e(\xi))=\left\{\begin{array}{ll} \xi, &  \xi\leq \xi_c=1/5\\
{\frac{-9\xi^2+10\xi-1}{16\xi}}, &\xi> \xi_c=1/5. \end{array}\right.
\end{array}\label{req1}
\end{equation}
Then one finds that $d^2{\cal E}_\xi(r_e(\xi))/d\xi^2$ is discontinuous at $\xi_c=1/5$ and the phase transition is said to
be of second order. For triatomic molecules, the (dimensionless) algebraic variational
coordinate $r$ has been related to a physical angular displacement \cite{Iachello3,curro,Iachello4}
(or `bending angle')  $\theta\sim r/a$, with $a$ the equilibrium bond length,
which reflects the degree of distortion of
the molecular framework from linearity ($r=0$). The displacement $r$ is also related to the coordinates
of P\"oschl-Teller (for the cylindrical oscillator) and Morse (for the displaced cylindrical oscillator)
potentials. In \eqref{energyns} we also see that $r$ provides a measure of the average vibrational
$\langle \hat n\rangle$ and squared angular momentum $\langle\hat{W}^2\rangle$ quantum numbers.

\begin{figure}
\includegraphics[width=8.5cm,angle=0]{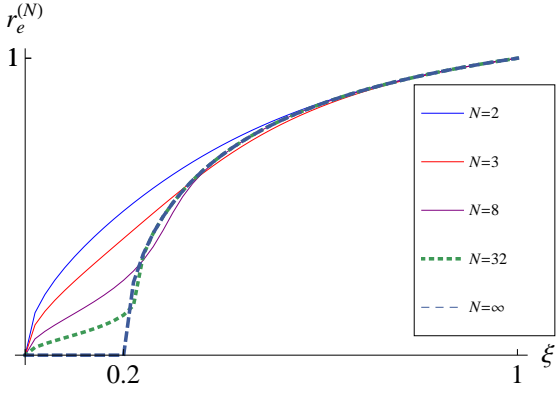}
\caption{Equilibrium radius $r_e^{(N)}(\xi)$ of even-cat states for $N=2, 3, 8, 32, \infty$, where we are identifying $r_e^{(\infty)}(\xi)=r_e(\xi)$.}\label{rop-s-ns}
\end{figure}

\begin{figure}
\includegraphics[width=8cm,angle=0]{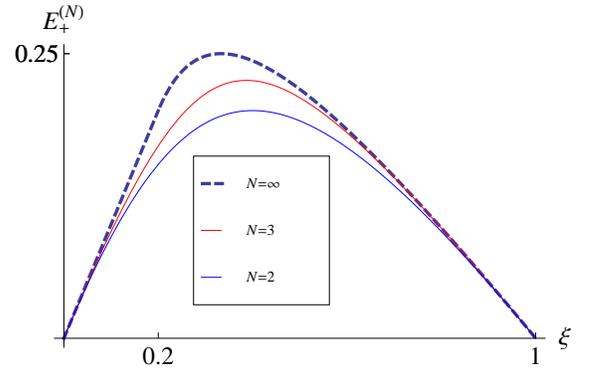}
\caption{Ground state energy per particle  $E^{(N)}_+(\xi)\equiv{\cal E}_{\xi+}^{(N)}(r_e^{(N)}(\xi))$ of even-cat states for
$N=2$, $N=3$ and $N=\infty$, where we are identifying ${\cal E}_{\xi+}^{(\infty)}(r_e^{(\infty)}(\xi))={\cal E}_\xi(r_e(\xi))$.}\label{energyN}
\end{figure}

Although  $|N;r_e(\xi)\rangle$ properly describes the ground state energy density 
in the thermodynamic limit $N\to\infty$,  we shall show that it does not capture the correct behavior for other ground state
properties sensitive to the parity symmetry $\hat\Pi$ of the Hamiltonian like,
for instance, vibration-rotation entanglement and delocalization measures. This is why we introduce in this article parity-symmetry-adapted CSs.
Indeed, a far better variational description
of the ground (resp. first-excited) state is given in terms of the even-(resp. odd)-parity projected `projective' CSs
\begin{equation}
|N;r,\pm\rangle\equiv\frac{(1\pm\hat\Pi)|N;r\rangle}{{\cal N}_\pm(r)}=
\frac{|N;r\rangle\pm|N;-r\rangle}{{\cal N}_\pm(r)},\label{even-pproj}
\end{equation}
where ${\cal N}_\pm(r)=\sqrt{2}(1\pm \langle N;-r|N;r\rangle)^{1/2}$ is a normalization constant, with
\begin{equation}
 \langle N;-r|N;r\rangle=((1-r^2)/(1+r^2))^N,\label{overlap}\end{equation}
and we have used that
$\hat{n}^k(\tau^\dag_x)^n|0\rangle=n^k(\tau^\dag_x)^n|0\rangle$, and therefore
$\hat\Pi(\tau^\dag_x)^n|0\rangle=(-\tau^\dag_x)^n|0\rangle$.
Note that the overlap $\langle N;-r|N;r\rangle$ is negligible for high $N$ (thermodynamic limit) and any $r>0$; therefore, in this limit, the
states (\ref{even-pproj}) are a superposition of two non-overlapping (distinguishable) quasi-classical (coherent) wave packets
(see \cite{casta1,casta2} and \cite{nuestro1,nuestro2,nuestro3} for a similar
behavior in the Dicke model). This justifies the term `Schr\"odinger catlike' for these states.
We shall only discuss the even (ground state ansatz) case here. Expanding $|N;r,+\rangle$ in the basis (\ref{basis}), as we did in (\ref{csinbasis}) 
for non-symmetric CSs, we
arrive to the new coefficients:
\begin{equation}
c_{n,m}^{(N,+)}(r)=\frac{c_{n,m}^{(N)}(r)+c_{n,m}^{(N)}(-r)}{{\cal N}_+(r)}.\label{evencoef}
\end{equation}
Note that now $c_{n,m}^{(N,+)}(r)=0$ for $n$=odd. The variational parameter $r$ is again computed by minimizing the ground state energy functional `per particle'
${\cal E}_{\xi,+}^{(N)}(r)=\langle\hat H\rangle_+/N$ as in (\ref{energyns}),
but now for the symmetric configuration $|N;r,+\rangle$,
given in terms of the new mean values:
\begin{eqnarray}
\frac{\langle\hat{n}\rangle_+}{N}&=&\frac{r^2((1+r^2)^{N-1}-(1-r^2)^{N-1})}{(1+r^2)^{N}+(1-r^2)^{N}}\\
\frac{\langle\hat{W}^2\rangle_+}{N}&=&2\frac{(1+r^2)^{N}+(1-r^2)^{N-2}(1+2Nr^2+r^4)}{(1+r^2)^{N}+(1-r^2)^{N}}.\nonumber
\end{eqnarray}
Unlike ${\cal E}_\xi(r)$,  the new energy functional ${\cal E}_{\xi,+}^{(N)}(r)$ depends on $N$.
From $\partial {\cal E}_{\xi,+}^{(N)}(r)/\partial r=0$ we can obtain the new equilibrium radius $r_e^{(N)}(\xi)$.
For example, for $N=2$ and $N=3$ we find analytic
explicit expressions for the equilibrium radius and the ground state energy per particle as a
function of the control parameter $\xi$:
\begin{equation}
 \begin{array}{rcl} r_e^{(2)}(\xi)&=&\sqrt{\frac{1}{2}-\frac{1}{2\xi}+\frac{\sqrt{1-2\xi+5\xi^2}}{2\xi}},\\
 {\cal E}_{\xi,+}^{(2)}(r_e^{(2)}(\xi))&=&\frac{1}{2}\left(1+\xi-\sqrt{1+\xi(-2+5\xi)}\right),\\
r_e^{(3)}(\xi)&=&\sqrt{\frac{2}{3}-\frac{1}{3\xi}+\frac{\sqrt{1-4\xi+7\xi^2}}{3\xi}},\\
 {\cal E}_{\xi,+}^{(3)}(r_e^{(3)}(\xi))&=&\frac{1}{3}\left(1+\xi-\sqrt{1+\xi(-4+7\xi)}\right).
\end{array}\label{req2}
\end{equation}
For higher values of $N$ we can compute $r_e^{(N)}(\xi)$ numerically.
Figure \ref{rop-s-ns} compares  $r_e(\xi)$ in (\ref{req1}) with $r_e^{(N)}(\xi)$ for
$N=2, 3, 8, 32$. We observe that the equilibrium radius $r_e^{(N)}(\xi)$ of the even-cat state $|N;r,+\rangle$ approaches the equilibrium radius $r_e(\xi)$ of the 
CS $|N;r\rangle$ in the thermodynamic limit, that is, $r_e^{(N)}(\xi)\to r_e(\xi)$ for $N\to\infty$. 
Moreover, one can also see (Figure \ref{energyN}) that the energy density ${\cal E}_{\xi+}^{(N)}(r_e^{(N)}(\xi))$ of the even-cat state 
is a increasing function of $N$ and that it approaches the energy density ${\cal E}_\xi(r_e(\xi))$ of the CS in the thermodynamic limit, that is, 
${\cal E}_{\xi,+}^{(N)}(r_e^{(N)}(\xi))\to {\cal E}_\xi(r_e(\xi))$ for $N\to\infty$ too. This behavior is displayed in Figure \ref{energyN} and it is also 
shared by the exact (numerical) density energy. Therefore, even-cat states 
provide a ground state energy description of the finite-size $(N<\infty)$ regime.
For future use, let us finish the description of variational states by providing the explicit expression of
the basis wave functions (\ref{basis}) in `position' $q_i=\frac{1}{\sqrt{2}}(a^\dag_i+a_i)$ representation, with
$(a_0,a_1,a_2)\equiv(\sigma,\tau_+,\tau_-)$ our three oscillator operators, in terms of Hermite polynomials $H_k(q)$:
\begin{equation}\begin{array}{rcl}
\langle q\,|N;n,l\rangle &=&
\frac{2^{-N/2}\pi^{-3/4}e^{-(q_0^2+q_1^2+q_2^2)/2}}{\sqrt{(N-n)!
\left(\frac{n+l}{2}\right)!\left(\frac{n-l}{2}\right)!}}\\ && \times
H_{N-n}(q_0)H_{\frac{n+l}{2}}(q_1)H_{\frac{n-l}{2}}(q_2).\end{array}\label{basis-posrep}
\end{equation}

\section{Vibration-rotation entanglement and delocalization measures\label{sec3}}
\begin{figure}
\includegraphics[width=8.5cm]{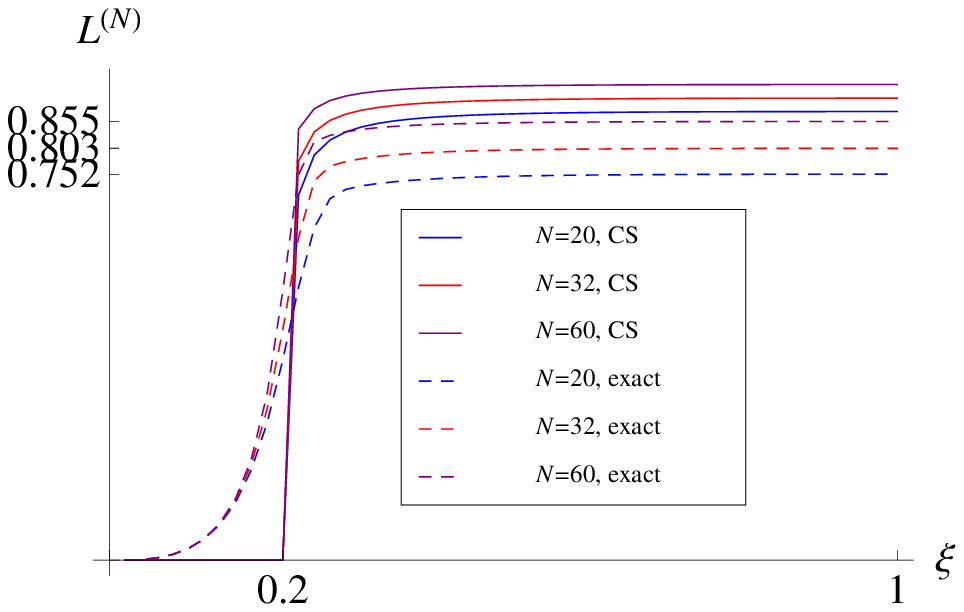}
\includegraphics[width=8.5cm]{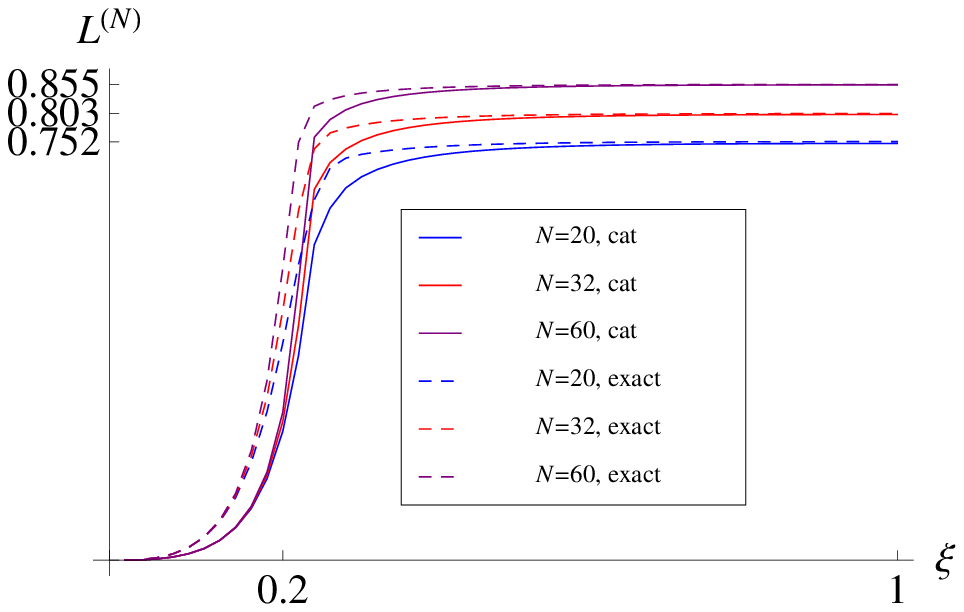}
\caption{Comparison of the exact linear entropy $L^{(N)}$ with the coherent state (top) and even-cat (bottom) 
variational approximations as a function of $\xi$ for $N=8,20,32$.}\label{entangN8-20}
\end{figure}

 Let us denote by
\begin{equation}
|\psi^{(N)}_\xi\rangle=\sum_{n=0}^N\sum_{m=0}^n c_{n,m}^{(N)}(\xi)|N;n,l=n-2m\rangle\label{exactw}
\end{equation}
the exact ground state of our system obtained by numerical diagonalization of the Hamiltonian (\ref{hamiltonian}) in terms of
the basis vectors (\ref{basis}) with coefficients $c_{n,m}^{(N)}(\xi)$, and by $\psi^{(N)}_\xi(q)=\langle q\,|N;r_e(\xi)\rangle$  the corresponding
wave function in position representation $q=(q_0,q_1,q_2)$, written in terms of Hermite polynomials \eqref{basis-posrep}. 
Let us consider the bipartite system given by vibrational $(q_0)$ and 2D-rotational $\vec{q}=(q_1,q_2)$ degrees of freedom.
Then, we can compute the reduced density matrix (RDM) for vibrational modes by integrating out
the rotational degrees of freedom:
\begin{equation}
 \rho^{(N)}_\xi(q_0,q'_0)=\int\int_{-\infty}^\infty d\vec{q}\, \psi^{(N)}_\xi(q_0,\vec{q})\bar{\psi}^{(N)}_\xi(q'_0,\vec{q}).
\end{equation}
The `purity' of  $\rho^{(N)}_\xi$ is given by:
\begin{eqnarray}
{\rm Tr}\left(\rho^{(N)}_\xi\right)^2&=&\int\int_{-\infty}^\infty dq_0dq_0' \rho^{(N)}_\xi(q_0,q'_0)\rho^{(N)}_\xi(q'_0,q_0)\nonumber\\
&=& \sum_{n=0}^N\left(\sum_{m=0}^n(c_{n,m}^{(N)}(\xi))^2\right)^2,\label{purity}
\end{eqnarray}
where we have used orthogonality properties of the basis vectors (\ref{basis}) and the coefficients of the expansion (\ref{exactw}).
Actually, the RDM $\rho^{(N)}_\xi$ is an $(N+1)\times (N+1)$ diagonal matrix:
\begin{equation}
\left(\rho^{(N)}_\xi\right)_{n,n'}=\lambda_n^{(N)}(\xi)\delta_{n,n'}
\end{equation}
with eigenvalues
\begin{equation}
\lambda_n^{(N)}(\xi)\equiv \sum_{m=0}^n (c_{n,m}^{(N)}(\xi))^2,
\end{equation}
and the vibrational quantum number $n$ running from $0$ to $N$. Using the coefficients \eqref{csinbasis} and \eqref{evencoef} for the
(non-symmetric) coherent state (CS)  and even-parity-adapted CS (cat) \eqref{pcs} and \eqref{even-pproj}, respectively, one can explicitly compute:
\begin{eqnarray}
 \lambda_n^{(N)}(\xi)_{\rm CS}&=&\binom{N}{n}\frac{r_e(\xi)^{2n}}{(1+r_e(\xi)^2)^N},\\
\lambda_n^{(N)}(\xi)_{\rm cat}&=&\binom{N}{n}\frac{(1+(-1)^n)r_e^{(N)}(\xi)^{2n}}{(1+r_e^{(N)}(\xi)^{2})^N+(1-r_e^{(N)}(\xi)^{2})^N}.\nonumber
\end{eqnarray}
The purity is then given in terms of hypergeometric functions as:
\begin{eqnarray}
 {\rm Tr}\left(\rho^{(N)}_\xi\right)^2_{\rm CS}&=&\frac{{}_2F_1(-N,-N;1;r_e(\xi)^4)}{(1+r_e(\xi)^2)^{2N}},\\
 {\rm Tr}\left(\rho^{(N)}_\xi\right)^2_{\rm cat}&=&2{\left((1+r_e^{(N)}(\xi)^{2})^N+(1-r_e^{(N)}(\xi)^{2})^N\right)^{-2}}\nonumber\\
&&\times\left({}_2F_1(-N,-N;1;r_e^{(N)}(\xi)^4)\right.\nonumber\\ &&+\left.{}_2F_1(-N,-N;1;-r_e^{(N)}(\xi)^4)\right).
\end{eqnarray}
Instead of ${\rm Tr}(\rho^{(N)}_\xi)^2$,
we shall use, for instance, the linear entropy:
\begin{equation}
 L^{(N)}(\xi)\equiv 1-{\rm Tr}\left(\rho^{(N)}_\xi\right)^2,\label{linentnosym}
\end{equation}
as a measure of entanglement. Since the size of $\rho^{(N)}_\xi$ is $N+1$, the linear entropy
$L^{(N)}(\xi)$ then ranges between $0$ (pure state)
and $N/(N+1)$ (completely mixed state).  Figure \ref{entangN8-20} compares the exact (numerical) linear entropy  with
that of the CS and cat states  (\ref{csinbasis}) and (\ref{even-pproj}), respectively.
One can see that linear entropy for the CS configuration gives wrong results in both phases.
Especially, in the second (bent) phase
for the maximal value at $\xi=1$, we have
\begin{equation}
 L^{(N)}_{\rm CS}(1)= 1-4^{1-N}\binom{2N}{N},\label{linentnosymxi1}
\end{equation}
which can be computed by taking into account that $r_e(1)=1$.
A much better agreement (remarkably for high $N$) is reached through the even-cat configuration, with linear entropy at $\xi=1$ given by:
\begin{equation}
 L^{(N)}_{\rm cat}(1)\simeq 1-2^{1-2N}\binom{2N}{N},\label{linentsymxi1}
\end{equation}
where we have also used that $r^{(N)}_e(1)=1,\forall N$.
Here we have made the approximation
\begin{equation}
\sum_{\nu=0}^{[N/2]}\binom{N}{2\nu}^2\simeq \frac{1}{2}\sum_{n=0}^{N}\binom{N}{n}^2=
\frac{1}{2}\binom{2N}{N}, \end{equation}
with $[N/2]=$Floor$(N/2)$, which is quite accurate even for relatively small values of $N$. Thus, we have that
the purity of the RDM for the even-cat is essentially half the purity for the CS at $\xi=1$;
a simple correction with important consequences that makes
(\ref{linentsymxi1}) a very good estimate for the entanglement linear entropy as a function of $N$
in the rigidly bent phase \cite{Iachello3}. Moreover, $L^{(N)}_{\rm cat}(\xi)= 1$ (maximal entanglement) for the rigidly bent phase ($\xi=1$) in the thermodynamic
limit $N=\infty$. 
In order to better evaluate the purity degree of our RDM for even-cat states, particularly 
for large $N$ and rigidly bent molecules, $\xi=1$,
we can use the Stirling's approximation
\begin{equation}
 {\rm Tr}\left(\rho^{(N)}_\xi\right)^2_{\rm cat}\simeq 2^{1-2N}\binom{2N}{N}\simeq \frac{2}{\sqrt{\pi N}},\;\;\;{\rm for}\;\;\;N\gg 1,
\end{equation}
which says that vibrational and rotational modes in the exact ground state (and in the even-cat variational approximation) are entangled 
but not maximally entangled in the bent phase for finite $N$, since the purity for a maximally entangled state is 
${\rm Tr}\left(\rho^{(N)}_\xi\right)^2_{\rm min.}=\frac{1}{N+1}$. 
In the `floppy region' \cite{Iachello3},  $0<\xi<\xi_c$, the linear entropy seems to converge to a non-zero value in the thermodynamic limit.
For rigidly linear molecules \cite{Iachello3}, $\xi=0$, the vibration-rotation entanglement linear entropy is zero.

\begin{figure}
\includegraphics[width=8.5cm]{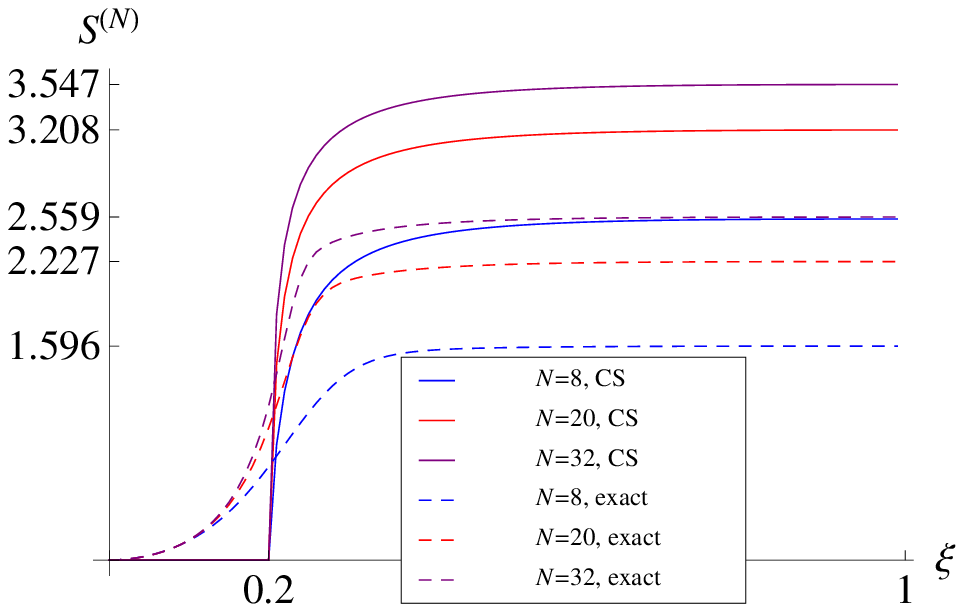}
\includegraphics[width=8.5cm]{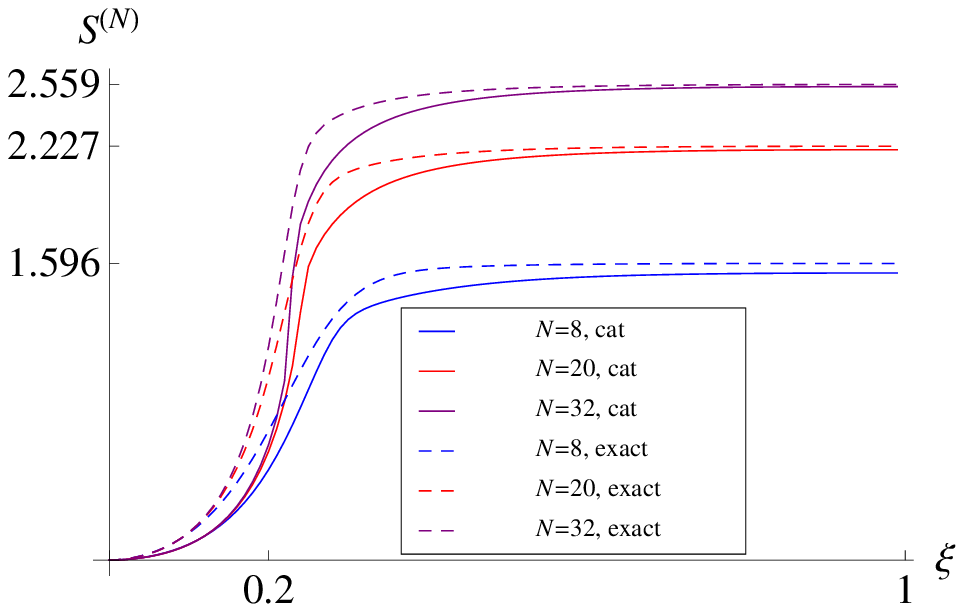}
\caption{Comparison of the exact von Neumann entropy $S^{(N)}$ with the coherent state (top) and even-cat (bottom)
variational approximations as a function of $\xi$ for $N=8,20,32$.}\label{vonneuN8-20}
\end{figure}

For completeness, we also provide in Figure \ref{vonneuN8-20} a comparative plot for the von Neumann entropy
\begin{eqnarray}
 S^{(N)}(\xi)&=&- {\rm Tr}\left(\rho^{(N)}_\xi\log_2 (\rho^{(N)}_\xi)\right)\nonumber\\ &=&-\sum_{n=0}^N \lambda_n^{(N)}(\xi)\log_2(\lambda_n^{(N)}(\xi)).
\end{eqnarray}
The qualitative  behavior of $S^{(N)}(\xi)$ is quite similar to that of $L^{(N)}(\xi)$. Actually, one can see again
that von Neumann entropy for the CS 
configuration gives wrong results in both phases. In particular, for high $N$ and rigidly bent molecules, $\xi=1$, von Neumann entropy behaves like:
\begin{eqnarray}
  S^{(N)}_{\rm cat}(1)&\simeq& \frac{1}{2} \log_2(N+1),\\ S^{(N)}_{\rm CS}(1)&\simeq& 1+\frac{1}{2} \log_2(N+1),\nonumber
\end{eqnarray}
denoting a von Neumann entropy excess of 1 of the CS with respect to the cat and exact configurations.

\begin{figure}
\includegraphics[width=8.5cm]{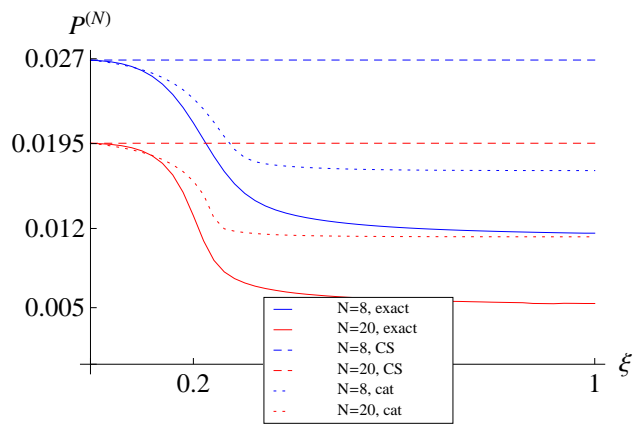}
\caption{Comparison of the exact inverse participation ratio $P^{(N)}$ with the coherent state (dashed-constant) and even-cat (dotted)
variational approximations as a function of $\xi$ for $N=8$ and $N=20$.}\label{ipr8-20}
\end{figure}

At this point one could still think that the CS approximation, albeit wrong for finite $N$,
still captures the essence of the growth of entanglement. Therefore, one could ask himself whether parity is
really essential to properly describe the ground state of vibron models in the thermodynamic limit or not.
To answer this question positively, we need to provide
a quantity from the ground state which is strongly sensitive to parity. This quantity turns out to be the `inverse participation ratio' (IPR):
\begin{equation}
P^{(N)}(\xi)= \int\int\int_{-\infty}^\infty dq_0dq_1dq_2\, |\psi^{(N)}_\xi(q_0,q_1,q_2)|^4.
\end{equation}
We can interpret the IPR as a measure of the
spread or delocalization of a wave function $\psi$ over a particular basis (here the position eigenfunctions $|q\rangle$),
much in the same way
the von Neumann entropy is a measure of the spread of a density
matrix $\rho$ over its diagonal basis. From Figure \ref{ipr8-20} we see that the exact (numerical) ground state wave function
exhibits a sudden delocalization across the phase transition, a spread that is also captured by the  even-cat \eqref{even-pproj} but goes unnoticed
in the case of the CS ansatz (\ref{pcs},\ref{csinbasis}), for which the IPR remains constant with the control parameter $\xi$.
This kind of behavior is also shared by the ground state in the Dicke model \cite{dicke},
where the wave packet in the normal phase splits up into two (almost) non-overlapping sub-packets in the super-radiant phase
(see e.g. \cite{casta1,casta2} and \cite{nuestro1,nuestro2,nuestro3}). Here too, the overlap \eqref{overlap} goes to zero 
for $\xi>\xi_c$ ($r>0$) in the thermodynamic limit $N\to \infty$, so that
the ground state wave function \eqref{even-pproj} is a linear superposition of two 
(almost) non-overlapping sub-packets in the bent phase.

For completeness, we also provide the IPR
\begin{equation}
\tilde P^{(N)}(\xi)= \sum_{n=0}^N\sum_{m=0}^n(c_{n,m}^{(N)}(\xi))^4,
\end{equation}
over the basis \eqref{basis} for the CS and cat ansatzs in terms of hypergeometric functions:
\begin{eqnarray}
\tilde P^{(N)}_{\rm CS}(\xi)&=&\frac{{}_3F_2(\frac{1}{2},-N,-N;1,1;r_e(\xi)^4)}{(1+r_e(\xi)^2)^{2N}},\\
\tilde P^{(N)}_{\rm cat}(\xi)&=&2{\left((1+r_e^{(N)}(\xi)^{2})^N+(1-r_e^{(N)}(\xi)^{2})^N\right)^{-2}}\nonumber\\
&&\times\left({}_3F_2(\frac{1}{2},-N,-N;1,1;r_e^{(N)}(\xi)^4)\right.\nonumber\\ &&+\left.{}_3F_2(\frac{1}{2},-N,-N;1,1;-r_e^{(N)}(\xi)^4)\right).
\end{eqnarray}

\section{Conclusions\label{sec4}}

We have obtained exact (numerical) results of entanglement 
and delocalization of the ground state in 2D vibron models for
finite-($N$)-size molecules. These two features provide sharp indicators of the shape QPT present in this model, denoting 
an abrupt change in the structure of the ground state at the critical point $\xi_c$. 

This result has been 
complemented and compared with two variational approximations (parity-symmetric and 
non-symmetric) which enrich the study. Even-parity (cat) configurations turn out to give a remarkably good variational description of 
entanglement measures (linear and von Neumann entropies), 
quantitatively reproducing the exact values for the entanglement entropy in the rigidly 
(linear and bent) phases and qualitatively capturing the entanglement 
entropy behavior in the ``floppy'' (intermediate) region. 

Results reveal the emergence of vibration-rotation
entanglement in the bent phase of vibron models, where vibrational and rotational modes
are entangled but not  maximally entangled. 

Unlike other ansatzs in the literature, these Schr\"odinger's
cat states do capture a delocalization of the ground state wave packet through the IPR $P^{(N)}(\xi)$ across the phase transition, thus
proving the relevance of parity symmetry for a proper description of the ground state in vibron models.

As a general comment, we know that entanglement is a crucial resource for information processing, being at the heart 
of quantum communication protocols and quantum computing efficient algorithms. Although the work we present here is theoretical, 
there arises the natural question  about an eventual experimental
feasibility of the obtained vibration-rotation
entanglement for quantum information processes. We know that phonon-roton scattering,
emission and absorption has been extensively studied in helium superfluid and we think that it is worth 
exploring this new possibility in molecules.

\section*{Acknowledgements.}  This work was supported by the Projects: FIS2011-24149 and FIS2011-29813-C02-01 
(Spanish MICINN), 
FQM-165/0207 and FQM219 (Junta de Andaluc\'\i a).
Discussions with F. P\'erez-Bernal are gratefully acknowledged.

\end{document}